\documentclass[english,superscriptaddress,prl,twocolumn]{revtex4}
\usepackage[toc,page,title,titletoc,header]{appendix}
\usepackage[T1]{fontenc}
\usepackage{color}
\usepackage{tikz}
\usepackage{enumerate}
\usepackage{bm}
\usepackage{graphicx}
\usepackage[latin1]{inputenc}
\usepackage{epstopdf}
\usepackage{amsthm}

\usepackage{amsmath}
\usepackage{shapepar}
\usepackage{color}

\usepackage{amssymb, amsbsy, amsthm}
\makeatletter

\newtheorem{Theorem}{Theorem}
\newtheorem{Lemma}{Lemma}

\usepackage{babel}
\makeatother
\begin{document}
\title{From  entanglement to discord: a perspective based on  partial transposition}

\author{Liang-Liang Sun\footnote[1]{email: sun18@ustc.edu.cn}}

\author{Xiang Zhou}

\author{Sixia Yu  \footnote[2]{email: yusixia@ustc.edu.cn}}
\affiliation{Department of Modern Physics and National Laboratory for Physical Sciences at Microscale, University of Science and Technology of China, Hefei, Anhui 230026, China}


\date{\today{}}
\begin{abstract}
 Here, we show that partial transposition, which is  initially  introduced to study entanglement, can also inspire  many results  on quantum discord including:  (I) a discord  criterion of spectrum invariant under partial transposition, stating that one state must contain discord if its spectrum is changed by  the action of partial transposition, (II)  an approach to  estimate  the geometric quantum discord and  the one-way deficit based on the change of  spectrum. To compare with entanglement theory,  we also  lower-bound the geometric quantum entanglement and the entanglement  of relative entropy. Thus, on one hand,  we illustrate an approach  to specify and estimate discord based on  partial transposition.  On the other hand,   we show that,  entanglement and discord, two basic notions of nonclassical correlations, can be  placed on the same ground such that their interplay and distinction  can be  illustrated in within a universal framework..
 \end{abstract}


\maketitle
\section{I. Introduction}
One distinctive feature of quantum theory is that  quantum systems can demonstrate various forms of nonclassical correlations, which  often find applications in quantum information science~\cite{RevModPhys.81.865, RevModPhys.86.419,  RevModPhys.84.1655, Bera_2018}.   Entanglement was the first such notion to be known about and then  appreciated as a key resource in  many quantum information tasks, such as quantum teleportation~\cite{PhysRevLett.70.1895},  cryptography~\cite{PhysRevLett.67.661}, quantum
algorithms~\cite{PhysRevLett.86.5188, PhysRevA.68.022312, Briegel2009},  metrology~\cite{Giovannetti2004}. Entanglement was  even believed to be  responsible for why   quantum resources can outperform  classical ones.  This belief  started to change when quantum discord is discovered, which  goes beyond entanglement and exists in a wide range of  quantum states that may be separable.  Like entanglement, discord can also   demonstrate quantum advantages  in diversified tasks, such as mixed-state quantum computing~\cite{PhysRevLett.81.5672, PhysRevLett.100.050502},   bounding distributed entanglement~\cite{PhysRevLett.109.070501}, remote state preparation~\cite{Daki__2012}, and  quantum state merging~\cite{PhysRevA.83.032324, PhysRevA.83.032323}. Although it is known  that  entanglement must demonstrate  quantum discord,   the specification and quantification  of these two basic  properties are commonly seen as  distinct subjects and follow  different lines of research.
 One interesting question is whether or not  these two properties can be characterized and quantified   within a universal framework.

Much focus has been  placed on  entanglement and  many powerful  tools are introduced~\cite{RevModPhys.81.865, Friis_2018}.   One may concern   whether  these tools  can be borrowed to  specify and estimate the relatively less studied discord. For this purpose, we consider the primary tool of of detecting   entanglement, namely,   the positive partial transpose (PPT) criterion~\cite{PhysRevLett.77.1413, HORODECKI19961},  with which,  one state is certified to be entangled if its transposed density matrix presents  negative eigenvalue.  This criterion  also induces one   computable  entanglement quantifier, referred to as    negativity~\cite{eisert2006entanglement, PhysRevA.65.032314},  as the  sum of   negative eigenvalues in absolute values.  Although the tool of PPT was related to  discord, $e.g.,$  in $2\times n$-dimensional system,   all discord-free states belong to a subclass of PPT states,  called  strong PPT states~\cite{PhysRevA.77.022113},  quantum discord is mainly detected  with discord witness~\cite{PhysRevA.84.032122, yu2011witnessing}.    To quantify  discord, quite a few measures have been introduced~\cite{RevModPhys.84.1655, Bera_2018}. Unfortunately, most of them  are hard to compute.  One exception is  the geometric quantum  discord (GQD)~\cite{PhysRevLett.105.190502, PhysRevLett.106.120401, PhysRevA.82.034302}, which   can be equivalently defined as the minimal disturbance under a projective measurement performed on one partite thus can be computed via an optimization on the measurement~\cite{PhysRevLett.106.120401, PhysRevA.82.034302, PhysRevA.86.042123,PhysRevA.85.024102, PhysRevA.85.024302}.

In this paper, we show  that  many results on quantum discord  can be inspired by the map of partial transposition.   We first illustrate that,  the spectrum  of  discord-free states are invariant (up to a relabel of indexes) under partial transposition. Therefore,  the change of spectrum  indicates the presence of discord.  We refer this criterion to as spectrum invariant under partial transposition (SIPT)  in contrast to the PPT in entanglement theory.  Quantitatively, we find that  the  spectrum change implies a lower bound on the GQD and a discord-like quantity of  one-way deficit. For the sake of comparing discord and entanglement,  we also provide lower bounds  for  the geometric quantum entanglement and the entanglement  of relative entropy. Thus, a perspective based on partial transposition is provided, in which,  entanglement is specified and estimated based on the negative eigenvalues presented under the map, in contrast, discord is specified and estimated based on the change of the spectrum.


\section{II.Preliminaries}
Let us first make a brief review on some  basic notions.
One quantum state $\rho_{AB}$ is said to be entangled if it cannot  be written as  the mixture of product states, namely,
 \begin{eqnarray}
\rho_{AB}=\textstyle \sum_{i}f_{i}\cdot \rho_{i, A}\otimes\rho_{i, B}. \label{sep}
\end{eqnarray}
One can expand  a general bipartite state in $N\otimes M$ dimensional Hilbert's space, entangled or separable,   in a chosen product basis as
\begin{eqnarray}
\textstyle \rho_{AB}=\sum_{i,j}^{N}\sum_{k,l}^{M}\rho_{ij, kl} |i\rangle \langle j|_{A}\cdot |k\rangle \langle l|_{B}.
\end{eqnarray}
 Henceforth we shall omit the footnote $A,B$ where not introducing ambiguity.
Given this decomposition, performing a partial  transposition on Alice's side leads to
 \begin{eqnarray}
\rho^{\Gamma_{A}}=\sum_{i,j}^{N}\sum_{k,l}^{M}\rho_{ij, kl} |j\rangle \langle i|\cdot |k\rangle \langle l|.
\end{eqnarray}
The density matrix of any separable state is positive, and one state is certified to be entangled if  it violates PPT criterion.  PPT
 criterion also induces  an entanglement quantifier of negativity  as  the sum of the negative spectrum in the absolute value as
\begin{eqnarray}
\textstyle \mathcal{N}:=\frac{\|\rho^{\Gamma_{A}}\|_{\rm tr}-1}{2},
\end{eqnarray}
where $\|A\|_{\rm tr}={\rm Tr}\sqrt{A A^{\dagger}}$.

 Quantum discord (QD) is a typical quantum correlation referring to the phenomenon that a composite quantum system contains more information than the subsystems taken separably. By definition, one discord-free state $\rho$  can be written in the form of
 \begin{eqnarray}
\rho=\sum_{i}f_{i}\cdot|i\rangle \langle i|\otimes\rho_{i}.
\end{eqnarray}
where $\{|i\rangle \}$ is one set of orthogonal basis.
It is clear that an entangled state must contain discord, however, it is not necessary for the converse.

\section{III. The criterion of spectrum invariant partial transposition}
The operation  of partial transposition    depends on the partite and basis  on which it is performed. However, the spectrum  of the partially  transposed state   is  independent on them.  We specify the spectrum of $\rho$  by $\lambda^{\downarrow}(\rho)$ shorten by $\lambda^{\downarrow}$, where the elements are   arranged  in  descending order, namely, $\lambda^{\downarrow}=(\lambda^{\downarrow}_{1}, \lambda^{\downarrow}_{2}, \cdots \lambda^{\downarrow}_{M\cdot N})$ with $\lambda^{\downarrow}_{i}\geq \lambda^{\downarrow}_{i+1}$, $\forall i$.    The fact that  $\lambda^{\downarrow}(\rho^{\Gamma_{A}})=
\lambda^{\downarrow}(\rho^{\Gamma_{B}})$
  can be illustrated via $\lambda^{\downarrow}(\rho^{\Gamma_{A}})=\lambda^{\downarrow} ([\rho^{\Gamma_{A}}]^{\Gamma})=\lambda^{\downarrow}(\rho^{\Gamma_{B}})$, where $\Gamma$ specifies the usual transposition acting on the joint system, which  does not alter  state's spectrum. The different choices of basis  to perform partial transposition can be captured with  a local unitary operation ${\rm U}_{A}$ acting  on Alice's side.  The  basis-independence  feature  can be illustrated as  $\lambda^{\downarrow}[(\operatorname{U}_{A}\rho \operatorname{U}^{\dagger}_{A})^{\Gamma_{A}}]=\lambda^{\downarrow}[(\operatorname{U}_{A}\rho \operatorname{U}^{\dagger}_{A})^{\Gamma_{B}}]=\lambda^{\downarrow}[\operatorname{U}_{A}(\rho^{\Gamma_{B}}) \operatorname{U}^{\dagger}_{A}]= \lambda^{\downarrow}[\rho^{\Gamma_{B}}]= \lambda^{\downarrow}[\rho^{\Gamma_{A}}]$.  Therefore, the spectrum of partially transposed state  is determined solely by state $\rho$.

Consider a discord-free state $\rho=\sum_{i}f_{i}\cdot|i\rangle \langle i|_{A}\otimes\rho_{i, B}$,  one can perform  partial transposition with respect to basis  $\{|i\rangle_{A}\}$ without losing  any generality. Definitely,  the state is invariant under the operation. An immediately consequence is our criterion of  SIPT
 \begin{Theorem}[Spectrum invariant under partial transposition]
Spectrum of a discord-free state $\rho$  is invariant under partial transposition:
 \begin{eqnarray}
\lambda^{\downarrow}(\rho)=\lambda^{\downarrow}(\rho^{\Gamma_{A}}).
\end{eqnarray}
\end{Theorem}
Thus, from the perspective of partial transposition,  a change  of spectrum  under the map indicates a non-trivial discord.   To certify quantum entanglement,     the change needs  to be large enough to ensure the presence of negative spectrum.

To justify discord via Theorem.1,  one can  compute the spectrum of $\rho$ and $\rho^{\Gamma_{A}}$, then makes a statement after comparing them. Or alteratively, using the method based on the moments of  matrix defined as $\Pi_{n}(\rho):={\rm Tr}(\rho^{n})$. Noting that,
 \begin{eqnarray}
\Pi_{n}(\rho)=\Pi_{n}(\rho^{\Gamma_{A}}), \forall n, if,  \lambda^{\downarrow}(\rho)=\lambda^{\downarrow}(\rho^{\Gamma_{A}}). \label{7}
\end{eqnarray}
 Such a justification begins  with the third moment as $\Pi_{1}(\rho)=\Pi_{1}(\rho^{\Gamma_{A}})$  and $\Pi_{2}(\rho)=\Pi_{2}(\rho^{\Gamma_{A}})$  are trivially satisfied, which is due to the fact partial transposition is trace-preserving and $\Pi_{2}(\rho)=\Pi_{2}(\rho^{\Gamma_{A}})$ is trivially satisfied. To deal with   a state $\rho$ of $M\otimes N-$dimensional system, a sufficient judgement of Eq.(\ref{7})  stops at most in the  $(M\cdot N+2)-$th moment.

 It is interesting to ask a question whether  SIPT provides a sufficient and necessary justification of discord.   Unfortunately, it is not the case as there are states containing discord while the spectrum is invariant under partial transposition, such as $X$ state having equal anti-diagonal terms~\cite{PhysRevA.81.042105}. This  is  similar to the case in entanglement theory, where  there are bounded entangled states standing positivity under partial transposition, for which, PPT criterion fails.



\section{IV. Quantitative estimation of discord and discord-like quantity}
Motivated by the fact that the negative spectrum of a partially transposed
 state can  be used to quantify entanglement,  we  consider   how the spectrum change under partial transposition quantitatively  relates to quantum discord.

\subsection{A. Lower-bound GQD}
GQD is one of the most studied discord measures  and defined as the minimum Hilbert-Schmidt  distance or $2-$norm between state of interest $\rho$ and  the set of  discord-free states specified by  $\mathcal{D}$~\cite{PhysRevLett.105.190502, PhysRevLett.106.120401, PhysRevA.82.034302}
 \begin{eqnarray}
D_{\rm HS}=\min_{\varrho\in \mathcal{D}}\|\rho-\varrho\|^{2}_{2}.
\end{eqnarray}
Now,  we  provide two analytic lower bounds on GQD.
We first make a quite useful observation:   $2-$norm is  invariant under partial transposition as it only involves the sum of the square   modulus  of matrix elements which is  invariant under partial transposition.  This observation implies   $\|\rho-\varrho\|^{2}_{2}=\|\rho^{\Gamma_{A}}-\varrho^{\Gamma_{A}}\|^{2}_{2}$. For  latter use, we provide an estimate of the quantity $\min_{\lambda_{d}}\|\lambda'^{\downarrow}-\lambda^{\downarrow}_{d}\|^{2}_{2}$ with  ${\lambda'}^{\downarrow}$ specifying the spectrum of $\rho^{\Gamma_{A}}$, and a general positive and normalized spectrum is specified  by $\lambda_{d}$
 with  $\lambda^{\downarrow}_{d, i} \geq 0$,  $\forall$ $i$ and $\sum_{i}\lambda^{\downarrow}_{d, i}=1$.
\begin{Lemma}
An analytic lower bound  on $\min_{\lambda_{d}}\|{\lambda'}^{\downarrow}-\lambda^{\downarrow}_{d}\|^{2}_{2}$ is given  as (see Supplemental Materials (SM) for proof)
 \begin{eqnarray}
\min_{\lambda_{d}}\|{\lambda'}^{\downarrow}-\lambda^{\downarrow}_{d}\|^{2}_{2} \geq \|{\lambda'}_{(2)}\|^{2}_{2}+ n\tau^{2}:=\mathfrak{L}({\lambda'}^{\downarrow}),
\end{eqnarray}
where we have partitioned ${\lambda'}^{\downarrow}$ into two parts, namely,  ${\lambda'}^{\downarrow}={\lambda'}^{\downarrow}_{(1)}\bigcup{\lambda'}^{\downarrow}_{(2)}$ with ${\lambda'}^{\downarrow}_{(1)}=\{{\lambda'}^{\downarrow}_{1}, \cdots, {\lambda'}^{\downarrow}_{n}\}$ and ${\lambda'}^{\downarrow}_{(2)}=\{{\lambda'}^{\downarrow}_{n+1}, \cdots, {\lambda'}^{\downarrow}_{MN}\}$ with $n$ specifying  the minimal  number such that $\tau:=\frac{{\sum^{n}_{i=1}{\lambda'}^{\downarrow}_{i}}-
1}{n}\geq{\lambda'}^{\downarrow}_{n+1}$ and $MN$ are the number of eigenvalues. The $\lambda_{d}$ achieving the minimum is specified by $\lambda_{\rm PPT}$
\end{Lemma}

Consider that  a discord-free state is  PPT, we have a lower-bound on  GQD as
 \begin{eqnarray}
D_{\rm HS}&=&\min_{\varrho\in \mathcal{D}}\|\rho-\varrho\|^{2}_{2}=
\|\rho^{\Gamma_{A}}-\varrho^{\Gamma_{A}}_{\min}\|^{2}_{2}\nonumber \\
&\geq&\|{\lambda'}^{\downarrow}-\lambda^{\downarrow}_{\min} \|^{2}_{2} \nonumber \\
&\geq& \textstyle \min_{\lambda^{\downarrow}_{d}} \|{\lambda'}^{\downarrow}-\lambda^{\downarrow}_{d}\|^{2}_{2}\nonumber \\
&=&
 \textstyle \mathfrak{L}(\lambda'):= L^{D}_{\rm PPT}, \label{lppt1} \\
&\geq & \textstyle \frac{\mathcal{N}^{2}}{N_{+}}+\frac{\mathcal{N}^{2}}{N_{-}}:={L'}^{D}_{\rm PPT},  \label{lppt2}
\end{eqnarray}
where we have specified   the closest discord-free state to $\rho$ (with respect to Hilbert-Schmidt distance) by $\rho_{\min}$ and  $N_{+}$ and $N_{-}$ are the numbers of positive and negative elements of $\lambda'$, respectively.  The first  inequality follows from   the inequality  that, for any two states $\rho$ and $\varrho$, it holds  $\|\rho-\varrho\|^{2}_{2}\geq \|{\lambda'}^{\downarrow}(\rho)-\lambda^{\downarrow}(\varrho)\|^{2}_{2}$~\cite{article1}. Eq.(\ref{lppt1}) is due to the  above Lemma, and  the proof for  Eq.(\ref{lppt2}) is left in SM.

Note that the lower bounds ${L}^{D}_{\rm PPT}$ and ${L'}^{D}_{\rm PPT}$ are non-zero only if $\rho$ violates PPT.   For a general case, one state containing discord may not violate PPT criterion, namely, the corresponding transposed density matrix is positive. Here,  we  give another lower bound, which may  be  non-trivial for such states.  Based on  the fact that any discord-free state is SIPT, we have
\begin{eqnarray}
2D_{\rm HS}&=&\min_{\varrho\in \mathcal{D}}\{\|\rho-\varrho\|^{2}_{2}+\|\rho^{\Gamma_{A}}-\varrho^{\Gamma_{A}}\|^{2}_{2}\nonumber \}\\
&\geq&\|\lambda^{\downarrow}-\lambda^{\downarrow}(\rho_{\min})\|^{2}_{2}+
\|{\lambda'}^{\downarrow}-\lambda^{\downarrow}(\rho_{\min})\|^{2}_{2}\nonumber \\
&=& \textstyle \sum_{i}[  {\lambda_{i}^{\downarrow}}^{2} +{{\lambda'}_{i}^{\downarrow}}^{2} +2{\lambda_{i, \min}^{\downarrow}}^{2}-2(\lambda_{i}^{\downarrow}+
{\lambda'}_{i}^{\downarrow})
\lambda_{i, \min}^{\downarrow}]\nonumber\\
&= &  \textstyle \sum_{i} [ {\lambda_{i}^{\downarrow}}^{2} +{{\lambda'}_{i}^{\downarrow}}^{2} -\frac{({{\lambda'}_{i}^{\downarrow}}+ {\lambda_{i}^{\downarrow}})^{2}}{2}+2(\frac{{{\lambda'}_{i}^{\downarrow}}+ {\lambda_{i}^{\downarrow}}}{2}-\lambda^{\downarrow}_{i\min})^{2}]\nonumber \\
&= &  \textstyle \frac{\|\lambda^{\downarrow}-{\lambda'}^{\downarrow}\|^{2}_{2}}{2}+ 2\|\frac{{{\lambda'}^{\downarrow}}+ {\lambda^{\downarrow}}}{2}-\lambda^{\downarrow}_{\min}\|^{2}_{2}\nonumber\\
&\geq&  \textstyle \frac{\|\lambda^{\downarrow}-{\lambda'}^{\downarrow}\|^{2}_{2}}{2}+ 2\min_{\lambda^{\downarrow}_{d}}\|\frac{{{\lambda'}^{\downarrow}}+ {\lambda^{\downarrow}}}{2}-\lambda^{\downarrow}_{d}\|^{2}_{2}\nonumber\\
&=&  \textstyle \frac{\|\lambda^{\downarrow}-{\lambda'}^{\downarrow}\|^{2}_{2}}{2}+\mathfrak{L}(\frac{{\lambda'}^{\downarrow}+{\lambda}^{\downarrow}}{2})\nonumber \\
&:=& 2L^{D}_{\rm SIPT}.
\end{eqnarray}
The $\lambda_{d}$ achieving the minimum is specified by $\lambda_{\rm SIPT}$.

 Note that the term $\|\lambda^{\downarrow}-{\lambda'}^{\downarrow}\|^{2}_{2}$ is non-zero as long as   state $\rho$ violates SIPT. Definitely, for PPT states, $L^{D}_{\rm SIPT}$ can be nontrivial while $L^{D}_{\rm PPT}=0$. One may concern whether $L^{D}_{\rm SIPT}\geq L^{D}_{\rm PPT}$.  It turns out to be  not the case.
 Although it is clear that $\|{\lambda'}^{\downarrow}-\lambda^{\downarrow}_{\rm PPT} \|^{2}_{2} \leq \|{\lambda'}^{\downarrow}-\lambda^{\downarrow}_{\rm SIPT} \|^{2}_{2}$ (where $\lambda^{\downarrow}_{\rm PPT}$ specifies the spectrum achieving  the minimum  of $\min_{\lambda_{d}}\|\lambda^{\downarrow}-\lambda^{\downarrow}_{d}\|^{2}_{2}$),  it is not necessary that $\|{\lambda}^{\downarrow}-\lambda^{\downarrow}_{\rm PPT} \|^{2}_{2}\leq \|{\lambda}^{\downarrow}-\lambda^{\downarrow}_{\rm SIPT} \|^{2}_{2}$. Consequently,  $L^{D}_{\rm PPT}$ can be larger than $L^{D}_{\rm SIPT}$. As an illustration, let us consider the maximum entangled state
$|\phi\rangle=\sum_{i}\frac{1}{\sqrt{d}}|ii\rangle$, for which, the exact value of $D_{\rm HS}=1-\frac{1}{d}$. By the above considerations,
we have $\lambda^{\downarrow}(|\phi\rangle\langle \phi|^{\Gamma_{A}})=\{{\lambda'}^{\downarrow}_{i}=\frac{1}{d};  1, \leq i \leq \frac{d(d+1)}{2}\}\cup \{ {\lambda'}^{\downarrow}_{i}=-\frac{1}{d}; \frac{d(d+1)}{2}< i \le d^{2}\}.$ We have $L^{D}_{\rm PPT}=1-\frac{2}{d+1}$ and
$L_{\rm SIPT}=\frac{3}{4}-\frac{1}{2d}-\frac{1}{2(d+1)}.  $ For maximal entangled state,  $L^{D}_{\rm PPT} > L^{D}_{\rm SIPT}$  when $d\geq 3$.  Given this fact,  we summarize our lower bounds on   $D_{\rm HS}$  as
\begin{Theorem}
The geometric quantum discord is lower-bounded as
 \begin{eqnarray}
 D_{\rm HS}\geq L^{\mathcal{D}}:=\max\{L^{D}_{\rm PPT}, L^{D}_{\rm SIPT}\}.
\end{eqnarray}
\end{Theorem}
 By this theorem, one can  estimate the GQD by the change of the spectrum of partially transposed  state  or the negative spectrum arising from the operation of partial transpose. In the following, we show that such a lower bound also induces the estimate of another discord-like quantity, namely,  one-way deficit.

\subsection{B. Lower-bound one-way deficit}
 Quantum work deficit~\cite{PhysRevLett.90.100402, Bera_2018} is introduced to capture  the connection between thermodynamics and information and defined as the additional extractable  information, or work from a bipartite quantum state when the two parties are in the same place as compared to the cases they are in  distant locations. Denote de-phasing operation on Bob's side as $ \{\Pi^{i}_{B}\}$ and $\varrho_{AB}=\sum_{i}{\rm I}\otimes \Pi^{i}_{B} \rho_{AB} {\rm I}\otimes \Pi^{i}_{B} $,    one-way deficit reads
 \begin{eqnarray}
\mathcal{WD}_{B}:=\min_{\{\Pi^{i}_{B}\}}S(\rho_{AB}\|\varrho_{AB}),
\end{eqnarray}
where $S(\rho\|\varrho):={\rm Tr}(\rho\ln \rho- \rho\ln \varrho)$ and  the minimization is taken  over all projective measurements performed on  Bob's side. Clearly, $\varrho_{AB}$ is a  discord-free state for any $ \{\Pi^{i}_{B}\}$.
We thus have
 \begin{eqnarray}
\min_{\{\Pi^{i}_{B}\}}S(\rho_{AB}\|\varrho_{AB})&\geq&
\min_{\varrho\in \Omega}S(\rho_{AB}\|\varrho)=S(\rho_{AB}\|\varrho_{\min,re})\nonumber\\
&\geq& \frac{\|\rho-\varrho_{\min,re}\|^{2}_{\rm tr}}{2\ln2}\geq\min_{\varrho'\in \Omega}\frac{\|\rho-\varrho'\|^{2}_{\rm tr}}{2\ln2}\nonumber\\
&=& \frac{\|\rho-\varrho_{\min, tr}\|^{2}_{\rm tr}}{2\ln2}\geq \frac{\|\rho-\varrho_{\min, tr}\|^{2}_{2}}{2\ln2} \nonumber\\
&\geq& \textstyle \frac{L^{\mathcal{D}}}{2\ln2}, \label{onedeficit}
\end{eqnarray}
where  $\varrho_{\min,re}$ specifies the closest  discord-free state of $\rho$ with respect to  the relative entropy, and $\rho_{\min, \rm tr}$ specifies the closest state  with respect to trace-distance, and  we have used the quantum Pinsker inequality  $S(\rho\| \varrho)\geq \textstyle \frac{1}{2\ln 2} \|\rho-\varrho\|^{2}_{\rm tr}$  in the second inequality and the norm inequality  $\|\rho-\varrho\|^{2}_{\rm tr}\geq  2\|\rho-\varrho\|^{2}_{2}$~\cite{PhysRevA.100.022103} in the fourth inequality.

\section{IV. Quantitative estimation of some entanglement measures}
Let us move to entanglement theory and estimate two entanglement measures that can be seen as the counterparts of the above ones in entanglement theory.

 The geometric of entanglement, which is defined as the minimal Hilbert-Schmidt distance between state of interest and the set of separable states specified by $\mathcal{S}$~\cite{Ozawa_2000} as
 \begin{eqnarray}
E_{\rm HS}:=\min_{\varrho\in \mathcal{S}}\|\rho-\varrho\|^{2}_{2}.
\end{eqnarray}
Note that  sparable states are PPT and $2-$norm is invariant under partial transposition. By the same consideration  in Eq.(\ref{lppt1}) as
 \begin{eqnarray}
E_{\rm HS}&=&\min_{\varrho\in \mathcal{S}}\|\rho^{\Gamma_{A}}-\varrho^{\Gamma_{A}}\|^{2}_{2}\nonumber \\
&\geq& L^{D}_{\rm PPT}\geq\textstyle \frac{\mathcal{N}^{2}}{N_{+}}+ \frac{\mathcal{N}^{2}}{N_{-}},\\
&\geq & \frac{4\mathcal{N}^{2}}{M\cdot N},
\end{eqnarray}
where we have used $N_{-}+N_{+}\leq M\cdot N$ in the third inequality. Our lower bound can be compared with the one reported in Ref~\cite{PhysRevA.86.024302}
$$E_{\rm HS}\geq \frac{\mathcal{N}^{2}}{\min\{(M-1)^{2}, (N-1)^{2}\}}. $$
Immediately, when $\frac{MN}{4}\leq \min\{(M-1)^{2}, (N-1)^{2}\}$, our lower bound is tighter. As  $N_{-}\leq (M-1)(N-1)$~\cite{PhysRevA.87.054301}, one  example in  case  is when $M=N$, $N_{-}\leq   (N-1)^{2}= \min\{(M-1)^{2}, (N-1)^{2}\}$ follows.


For a comparison to the one-way deficit, we now consider the entanglement of relative entropy.
This measure serves as an upper bound for the entanglement of distillation, namely, the  minimal number of singlets that are needed to build a single copy of the concerned state. By the similar derivation done in Eq.(\ref{onedeficit}), we have
 \begin{eqnarray}
E_{\rm re}:=\min_{\varrho\in \mathcal{S}}S(\rho\|\varrho)\geq \frac{L^{D}_{\rm PPT}}{2\ln 2}.
\end{eqnarray}

\emph{Conclusion}
Entanglement and discord are two typical quantum correlations that are  generally studied separably. In this paper,    we show that one can use the primary tool of entanglement theory, namely, the map of partial transposition,  to study discord, and many aspects of the two notions can be connected  in one-to-one correspondence.   In contrast to the PPT in entanglement theory,  we show discord can be specified by the  change of spectrum of the density matrix after partial transposition,  which leads  to a discord criterion of SIPT.
 Analogously to the entanglement measure of negativity,   the change of spectrum is shown to  imply the estimation of the GQD and one-way deficit.  We also estimate the geometric quantum entanglement and the entanglement of relative entropy. In this way, we  provide not only one perspective to investigate discord but also, a hierarchical specification and quantitative estimation of entanglement and discord.

\bibliography{estimate}

\begin{thebibliography}{37}
\expandafter\ifx\csname natexlab\endcsname\relax\def\natexlab#1{#1}\fi
\expandafter\ifx\csname bibnamefont\endcsname\relax
  \def\bibnamefont#1{#1}\fi
\expandafter\ifx\csname bibfnamefont\endcsname\relax
  \def\bibfnamefont#1{#1}\fi
\expandafter\ifx\csname citenamefont\endcsname\relax
  \def\citenamefont#1{#1}\fi
\expandafter\ifx\csname url\endcsname\relax
  \def\url#1{\texttt{#1}}\fi
\expandafter\ifx\csname urlprefix\endcsname\relax\def\urlprefix{URL }\fi
\providecommand{\bibinfo}[2]{#2}
\providecommand{\eprint}[2][]{\url{#2}}

\bibitem[{\citenamefont{Horodecki et~al.}(2009)\citenamefont{Horodecki,
  Horodecki, Horodecki, and Horodecki}}]{RevModPhys.81.865}
\bibinfo{author}{\bibfnamefont{R.}~\bibnamefont{Horodecki}},
  \bibinfo{author}{\bibfnamefont{P.}~\bibnamefont{Horodecki}},
  \bibinfo{author}{\bibfnamefont{M.}~\bibnamefont{Horodecki}},
  \bibnamefont{and}
  \bibinfo{author}{\bibfnamefont{K.}~\bibnamefont{Horodecki}},
  \bibinfo{journal}{Rev. Mod. Phys.} \textbf{\bibinfo{volume}{81}},
  \bibinfo{pages}{865} (\bibinfo{year}{2009}),
  \urlprefix\url{https://link.aps.org/doi/10.1103/RevModPhys.81.865}.

\bibitem[{\citenamefont{Brunner et~al.}(2014)\citenamefont{Brunner, Cavalcanti,
  Pironio, Scarani, and Wehner}}]{RevModPhys.86.419}
\bibinfo{author}{\bibfnamefont{N.}~\bibnamefont{Brunner}},
  \bibinfo{author}{\bibfnamefont{D.}~\bibnamefont{Cavalcanti}},
  \bibinfo{author}{\bibfnamefont{S.}~\bibnamefont{Pironio}},
  \bibinfo{author}{\bibfnamefont{V.}~\bibnamefont{Scarani}}, \bibnamefont{and}
  \bibinfo{author}{\bibfnamefont{S.}~\bibnamefont{Wehner}},
  \bibinfo{journal}{Rev. Mod. Phys.} \textbf{\bibinfo{volume}{86}},
  \bibinfo{pages}{419} (\bibinfo{year}{2014}),
  \urlprefix\url{https://link.aps.org/doi/10.1103/RevModPhys.86.419}.

\bibitem[{\citenamefont{Modi et~al.}(2012)\citenamefont{Modi, Brodutch, Cable,
  Paterek, and Vedral}}]{RevModPhys.84.1655}
\bibinfo{author}{\bibfnamefont{K.}~\bibnamefont{Modi}},
  \bibinfo{author}{\bibfnamefont{A.}~\bibnamefont{Brodutch}},
  \bibinfo{author}{\bibfnamefont{H.}~\bibnamefont{Cable}},
  \bibinfo{author}{\bibfnamefont{T.}~\bibnamefont{Paterek}}, \bibnamefont{and}
  \bibinfo{author}{\bibfnamefont{V.}~\bibnamefont{Vedral}},
  \bibinfo{journal}{Rev. Mod. Phys.} \textbf{\bibinfo{volume}{84}},
  \bibinfo{pages}{1655} (\bibinfo{year}{2012}),
  \urlprefix\url{https://link.aps.org/doi/10.1103/RevModPhys.84.1655}.

\bibitem[{\citenamefont{Bera et~al.}(2017)\citenamefont{Bera, Das, Sadhukhan,
  Roy, Sen(De), and Sen}}]{Bera_2018}
\bibinfo{author}{\bibfnamefont{A.}~\bibnamefont{Bera}},
  \bibinfo{author}{\bibfnamefont{T.}~\bibnamefont{Das}},
  \bibinfo{author}{\bibfnamefont{D.}~\bibnamefont{Sadhukhan}},
  \bibinfo{author}{\bibfnamefont{S.~S.} \bibnamefont{Roy}},
  \bibinfo{author}{\bibfnamefont{A.}~\bibnamefont{Sen(De)}}, \bibnamefont{and}
  \bibinfo{author}{\bibfnamefont{U.}~\bibnamefont{Sen}}, \bibinfo{journal}{Rep.
  Prog. Phys.} \textbf{\bibinfo{volume}{81}}, \bibinfo{pages}{024001}
  (\bibinfo{year}{2017}),
  \urlprefix\url{https://dx.doi.org/10.1088/1361-6633/aa872f}.

\bibitem[{\citenamefont{Bennett et~al.}(1993)\citenamefont{Bennett, Brassard,
  Cr\'epeau, Jozsa, Peres, and Wootters}}]{PhysRevLett.70.1895}
\bibinfo{author}{\bibfnamefont{C.~H.} \bibnamefont{Bennett}},
  \bibinfo{author}{\bibfnamefont{G.}~\bibnamefont{Brassard}},
  \bibinfo{author}{\bibfnamefont{C.}~\bibnamefont{Cr\'epeau}},
  \bibinfo{author}{\bibfnamefont{R.}~\bibnamefont{Jozsa}},
  \bibinfo{author}{\bibfnamefont{A.}~\bibnamefont{Peres}}, \bibnamefont{and}
  \bibinfo{author}{\bibfnamefont{W.~K.} \bibnamefont{Wootters}},
  \bibinfo{journal}{Phys. Rev. Lett.} \textbf{\bibinfo{volume}{70}},
  \bibinfo{pages}{1895} (\bibinfo{year}{1993}),
  \urlprefix\url{https://link.aps.org/doi/10.1103/PhysRevLett.70.1895}.

\bibitem[{\citenamefont{Ekert}(1991)}]{PhysRevLett.67.661}
\bibinfo{author}{\bibfnamefont{A.~K.} \bibnamefont{Ekert}},
  \bibinfo{journal}{Phys. Rev. Lett.} \textbf{\bibinfo{volume}{67}},
  \bibinfo{pages}{661} (\bibinfo{year}{1991}),
  \urlprefix\url{https://link.aps.org/doi/10.1103/PhysRevLett.67.661}.

\bibitem[{\citenamefont{Raussendorf and Briegel}(2001)}]{PhysRevLett.86.5188}
\bibinfo{author}{\bibfnamefont{R.}~\bibnamefont{Raussendorf}} \bibnamefont{and}
  \bibinfo{author}{\bibfnamefont{H.~J.} \bibnamefont{Briegel}},
  \bibinfo{journal}{Phys. Rev. Lett.} \textbf{\bibinfo{volume}{86}},
  \bibinfo{pages}{5188} (\bibinfo{year}{2001}),
  \urlprefix\url{https://link.aps.org/doi/10.1103/PhysRevLett.86.5188}.

\bibitem[{\citenamefont{Raussendorf et~al.}(2003)\citenamefont{Raussendorf,
  Browne, and Briegel}}]{PhysRevA.68.022312}
\bibinfo{author}{\bibfnamefont{R.}~\bibnamefont{Raussendorf}},
  \bibinfo{author}{\bibfnamefont{D.~E.} \bibnamefont{Browne}},
  \bibnamefont{and} \bibinfo{author}{\bibfnamefont{H.~J.}
  \bibnamefont{Briegel}}, \bibinfo{journal}{Phys. Rev. A}
  \textbf{\bibinfo{volume}{68}}, \bibinfo{pages}{022312}
  (\bibinfo{year}{2003}),
  \urlprefix\url{https://link.aps.org/doi/10.1103/PhysRevA.68.022312}.

\bibitem[{\citenamefont{Briegel et~al.}(2009)\citenamefont{Briegel, Browne,
  D\"{u}r, Raussendorf, and den Nest}}]{Briegel2009}
\bibinfo{author}{\bibfnamefont{H.~J.} \bibnamefont{Briegel}},
  \bibinfo{author}{\bibfnamefont{D.~E.} \bibnamefont{Browne}},
  \bibinfo{author}{\bibfnamefont{W.}~\bibnamefont{D\"{u}r}},
  \bibinfo{author}{\bibfnamefont{R.}~\bibnamefont{Raussendorf}},
  \bibnamefont{and} \bibinfo{author}{\bibfnamefont{M.~V.} \bibnamefont{den
  Nest}}, \bibinfo{journal}{Nat. Phys} \textbf{\bibinfo{volume}{5}},
  \bibinfo{pages}{19} (\bibinfo{year}{2009}),
  \urlprefix\url{https://doi.org/10.1038%2Fnphys1157}.

\bibitem[{\citenamefont{Giovannetti et~al.}(2004)\citenamefont{Giovannetti,
  Lloyd, and Maccone}}]{Giovannetti2004}
\bibinfo{author}{\bibfnamefont{V.}~\bibnamefont{Giovannetti}},
  \bibinfo{author}{\bibfnamefont{S.}~\bibnamefont{Lloyd}}, \bibnamefont{and}
  \bibinfo{author}{\bibfnamefont{L.}~\bibnamefont{Maccone}},
  \bibinfo{journal}{Science} \textbf{\bibinfo{volume}{306}},
  \bibinfo{pages}{1330} (\bibinfo{year}{2004}),
  \urlprefix\url{https://doi.org/10.1126%2Fscience.1104149}.

\bibitem[{\citenamefont{Knill and Laflamme}(1998)}]{PhysRevLett.81.5672}
\bibinfo{author}{\bibfnamefont{E.}~\bibnamefont{Knill}} \bibnamefont{and}
  \bibinfo{author}{\bibfnamefont{R.}~\bibnamefont{Laflamme}},
  \bibinfo{journal}{Phys. Rev. Lett.} \textbf{\bibinfo{volume}{81}},
  \bibinfo{pages}{5672} (\bibinfo{year}{1998}),
  \urlprefix\url{https://link.aps.org/doi/10.1103/PhysRevLett.81.5672}.

\bibitem[{\citenamefont{Datta et~al.}(2008)\citenamefont{Datta, Shaji, and
  Caves}}]{PhysRevLett.100.050502}
\bibinfo{author}{\bibfnamefont{A.}~\bibnamefont{Datta}},
  \bibinfo{author}{\bibfnamefont{A.}~\bibnamefont{Shaji}}, \bibnamefont{and}
  \bibinfo{author}{\bibfnamefont{C.~M.} \bibnamefont{Caves}},
  \bibinfo{journal}{Phys. Rev. Lett.} \textbf{\bibinfo{volume}{100}},
  \bibinfo{pages}{050502} (\bibinfo{year}{2008}),
  \urlprefix\url{https://link.aps.org/doi/10.1103/PhysRevLett.100.050502}.

\bibitem[{\citenamefont{Chuan et~al.}(2012)\citenamefont{Chuan, Maillard, Modi,
  Paterek, Paternostro, and Piani}}]{PhysRevLett.109.070501}
\bibinfo{author}{\bibfnamefont{T.~K.} \bibnamefont{Chuan}},
  \bibinfo{author}{\bibfnamefont{J.}~\bibnamefont{Maillard}},
  \bibinfo{author}{\bibfnamefont{K.}~\bibnamefont{Modi}},
  \bibinfo{author}{\bibfnamefont{T.}~\bibnamefont{Paterek}},
  \bibinfo{author}{\bibfnamefont{M.}~\bibnamefont{Paternostro}},
  \bibnamefont{and} \bibinfo{author}{\bibfnamefont{M.}~\bibnamefont{Piani}},
  \bibinfo{journal}{Phys. Rev. Lett.} \textbf{\bibinfo{volume}{109}},
  \bibinfo{pages}{070501} (\bibinfo{year}{2012}),
  \urlprefix\url{https://link.aps.org/doi/10.1103/PhysRevLett.109.070501}.

\bibitem[{\citenamefont{Daki{\'{c}} et~al.}(2012)\citenamefont{Daki{\'{c}},
  Lipp, Ma, Ringbauer, Kropatschek, Barz, Paterek, Vedral, Zeilinger, Brukner
  et~al.}}]{Daki__2012}
\bibinfo{author}{\bibfnamefont{B.}~\bibnamefont{Daki{\'{c}}}},
  \bibinfo{author}{\bibfnamefont{Y.~O.} \bibnamefont{Lipp}},
  \bibinfo{author}{\bibfnamefont{X.}~\bibnamefont{Ma}},
  \bibinfo{author}{\bibfnamefont{M.}~\bibnamefont{Ringbauer}},
  \bibinfo{author}{\bibfnamefont{S.}~\bibnamefont{Kropatschek}},
  \bibinfo{author}{\bibfnamefont{S.}~\bibnamefont{Barz}},
  \bibinfo{author}{\bibfnamefont{T.}~\bibnamefont{Paterek}},
  \bibinfo{author}{\bibfnamefont{V.}~\bibnamefont{Vedral}},
  \bibinfo{author}{\bibfnamefont{A.}~\bibnamefont{Zeilinger}},
  \bibinfo{author}{\bibfnamefont{{\v{C}}.}~\bibnamefont{Brukner}},
  \bibnamefont{et~al.}, \bibinfo{journal}{Nat. Phys}
  \textbf{\bibinfo{volume}{8}}, \bibinfo{pages}{666} (\bibinfo{year}{2012}),
  \urlprefix\url{https://doi.org/10.1038%2Fnphys2377}.

\bibitem[{\citenamefont{Cavalcanti et~al.}(2011)\citenamefont{Cavalcanti,
  Aolita, Boixo, Modi, Piani, and Winter}}]{PhysRevA.83.032324}
\bibinfo{author}{\bibfnamefont{D.}~\bibnamefont{Cavalcanti}},
  \bibinfo{author}{\bibfnamefont{L.}~\bibnamefont{Aolita}},
  \bibinfo{author}{\bibfnamefont{S.}~\bibnamefont{Boixo}},
  \bibinfo{author}{\bibfnamefont{K.}~\bibnamefont{Modi}},
  \bibinfo{author}{\bibfnamefont{M.}~\bibnamefont{Piani}}, \bibnamefont{and}
  \bibinfo{author}{\bibfnamefont{A.}~\bibnamefont{Winter}},
  \bibinfo{journal}{Phys. Rev. A} \textbf{\bibinfo{volume}{83}},
  \bibinfo{pages}{032324} (\bibinfo{year}{2011}),
  \urlprefix\url{https://link.aps.org/doi/10.1103/PhysRevA.83.032324}.

\bibitem[{\citenamefont{Madhok and Datta}(2011)}]{PhysRevA.83.032323}
\bibinfo{author}{\bibfnamefont{V.}~\bibnamefont{Madhok}} \bibnamefont{and}
  \bibinfo{author}{\bibfnamefont{A.}~\bibnamefont{Datta}},
  \bibinfo{journal}{Phys. Rev. A} \textbf{\bibinfo{volume}{83}},
  \bibinfo{pages}{032323} (\bibinfo{year}{2011}),
  \urlprefix\url{https://link.aps.org/doi/10.1103/PhysRevA.83.032323}.

\bibitem[{\citenamefont{Friis et~al.}(2018)\citenamefont{Friis, Vitagliano,
  Malik, and Huber}}]{Friis_2018}
\bibinfo{author}{\bibfnamefont{N.}~\bibnamefont{Friis}},
  \bibinfo{author}{\bibfnamefont{G.}~\bibnamefont{Vitagliano}},
  \bibinfo{author}{\bibfnamefont{M.}~\bibnamefont{Malik}}, \bibnamefont{and}
  \bibinfo{author}{\bibfnamefont{M.}~\bibnamefont{Huber}},
  \bibinfo{journal}{Nat. Rev. Phys.} \textbf{\bibinfo{volume}{1}},
  \bibinfo{pages}{72} (\bibinfo{year}{2018}),
  \urlprefix\url{https://doi.org/10.1038%2Fs42254-018-0003-5}.

\bibitem[{\citenamefont{Peres}(1996)}]{PhysRevLett.77.1413}
\bibinfo{author}{\bibfnamefont{A.}~\bibnamefont{Peres}},
  \bibinfo{journal}{Phys. Rev. Lett.} \textbf{\bibinfo{volume}{77}},
  \bibinfo{pages}{1413} (\bibinfo{year}{1996}),
  \urlprefix\url{https://link.aps.org/doi/10.1103/PhysRevLett.77.1413}.

\bibitem[{\citenamefont{Horodecki et~al.}(1996)\citenamefont{Horodecki,
  Horodecki, and Horodecki}}]{HORODECKI19961}
\bibinfo{author}{\bibfnamefont{M.}~\bibnamefont{Horodecki}},
  \bibinfo{author}{\bibfnamefont{P.}~\bibnamefont{Horodecki}},
  \bibnamefont{and}
  \bibinfo{author}{\bibfnamefont{R.}~\bibnamefont{Horodecki}},
  \bibinfo{journal}{Phys. Lett. A} \textbf{\bibinfo{volume}{223}},
  \bibinfo{pages}{1} (\bibinfo{year}{1996}), ISSN \bibinfo{issn}{0375-9601},
  \urlprefix\url{https://www.sciencedirect.com/science/article/pii/S0375960196007062}.

\bibitem[{\citenamefont{Eisert}(2006)}]{eisert2006entanglement}
\bibinfo{author}{\bibfnamefont{J.}~\bibnamefont{Eisert}},
  \emph{\bibinfo{title}{Entanglement in quantum information theory}}
  (\bibinfo{year}{2006}), \eprint{quant-ph/0610253}.

\bibitem[{\citenamefont{Vidal and Werner}(2002)}]{PhysRevA.65.032314}
\bibinfo{author}{\bibfnamefont{G.}~\bibnamefont{Vidal}} \bibnamefont{and}
  \bibinfo{author}{\bibfnamefont{R.~F.} \bibnamefont{Werner}},
  \bibinfo{journal}{Phys. Rev. A} \textbf{\bibinfo{volume}{65}},
  \bibinfo{pages}{032314} (\bibinfo{year}{2002}),
  \urlprefix\url{https://link.aps.org/doi/10.1103/PhysRevA.65.032314}.

\bibitem[{\citenamefont{Chru\ifmmode \acute{s}\else
  \'{s}\fi{}ci\ifmmode~\acute{n}\else \'{n}\fi{}ski
  et~al.}(2008)\citenamefont{Chru\ifmmode \acute{s}\else
  \'{s}\fi{}ci\ifmmode~\acute{n}\else \'{n}\fi{}ski, Jurkowski, and
  Kossakowski}}]{PhysRevA.77.022113}
\bibinfo{author}{\bibfnamefont{D.}~\bibnamefont{Chru\ifmmode \acute{s}\else
  \'{s}\fi{}ci\ifmmode~\acute{n}\else \'{n}\fi{}ski}},
  \bibinfo{author}{\bibfnamefont{J.}~\bibnamefont{Jurkowski}},
  \bibnamefont{and}
  \bibinfo{author}{\bibfnamefont{A.}~\bibnamefont{Kossakowski}},
  \bibinfo{journal}{Phys. Rev. A} \textbf{\bibinfo{volume}{77}},
  \bibinfo{pages}{022113} (\bibinfo{year}{2008}),
  \urlprefix\url{https://link.aps.org/doi/10.1103/PhysRevA.77.022113}.

\bibitem[{\citenamefont{Zhang et~al.}(2011)\citenamefont{Zhang, Yu, Chen, and
  Oh}}]{PhysRevA.84.032122}
\bibinfo{author}{\bibfnamefont{C.}~\bibnamefont{Zhang}},
  \bibinfo{author}{\bibfnamefont{S.}~\bibnamefont{Yu}},
  \bibinfo{author}{\bibfnamefont{Q.}~\bibnamefont{Chen}}, \bibnamefont{and}
  \bibinfo{author}{\bibfnamefont{C.~H.} \bibnamefont{Oh}},
  \bibinfo{journal}{Phys. Rev. A} \textbf{\bibinfo{volume}{84}},
  \bibinfo{pages}{032122} (\bibinfo{year}{2011}),
  \urlprefix\url{https://link.aps.org/doi/10.1103/PhysRevA.84.032122}.

\bibitem[{\citenamefont{Yu et~al.}(2011)\citenamefont{Yu, Zhang, Chen, and
  Oh}}]{yu2011witnessing}
\bibinfo{author}{\bibfnamefont{S.}~\bibnamefont{Yu}},
  \bibinfo{author}{\bibfnamefont{C.}~\bibnamefont{Zhang}},
  \bibinfo{author}{\bibfnamefont{Q.}~\bibnamefont{Chen}}, \bibnamefont{and}
  \bibinfo{author}{\bibfnamefont{C.~H.} \bibnamefont{Oh}},
  \emph{\bibinfo{title}{Witnessing the quantum discord of all the unknown
  states}} (\bibinfo{year}{2011}), \eprint{1102.4710}.

\bibitem[{\citenamefont{Daki\ifmmode~\acute{c}\else \'{c}\fi{}
  et~al.}(2010)\citenamefont{Daki\ifmmode~\acute{c}\else \'{c}\fi{}, Vedral,
  and Brukner}}]{PhysRevLett.105.190502}
\bibinfo{author}{\bibfnamefont{B.}~\bibnamefont{Daki\ifmmode~\acute{c}\else
  \'{c}\fi{}}}, \bibinfo{author}{\bibfnamefont{V.}~\bibnamefont{Vedral}},
  \bibnamefont{and} \bibinfo{author}{\bibfnamefont{i.~c.~v.}
  \bibnamefont{Brukner}}, \bibinfo{journal}{Phys. Rev. Lett.}
  \textbf{\bibinfo{volume}{105}}, \bibinfo{pages}{190502}
  (\bibinfo{year}{2010}),
  \urlprefix\url{https://link.aps.org/doi/10.1103/PhysRevLett.105.190502}.

\bibitem[{\citenamefont{Luo and Fu}(2011)}]{PhysRevLett.106.120401}
\bibinfo{author}{\bibfnamefont{S.}~\bibnamefont{Luo}} \bibnamefont{and}
  \bibinfo{author}{\bibfnamefont{S.}~\bibnamefont{Fu}}, \bibinfo{journal}{Phys.
  Rev. Lett.} \textbf{\bibinfo{volume}{106}}, \bibinfo{pages}{120401}
  (\bibinfo{year}{2011}),
  \urlprefix\url{https://link.aps.org/doi/10.1103/PhysRevLett.106.120401}.

\bibitem[{\citenamefont{Luo and Fu}(2010)}]{PhysRevA.82.034302}
\bibinfo{author}{\bibfnamefont{S.}~\bibnamefont{Luo}} \bibnamefont{and}
  \bibinfo{author}{\bibfnamefont{S.}~\bibnamefont{Fu}}, \bibinfo{journal}{Phys.
  Rev. A} \textbf{\bibinfo{volume}{82}}, \bibinfo{pages}{034302}
  (\bibinfo{year}{2010}),
  \urlprefix\url{https://link.aps.org/doi/10.1103/PhysRevA.82.034302}.

\bibitem[{\citenamefont{Miranowicz et~al.}(2012)\citenamefont{Miranowicz,
  Horodecki, Chhajlany, Tuziemski, and Sperling}}]{PhysRevA.86.042123}
\bibinfo{author}{\bibfnamefont{A.}~\bibnamefont{Miranowicz}},
  \bibinfo{author}{\bibfnamefont{P.}~\bibnamefont{Horodecki}},
  \bibinfo{author}{\bibfnamefont{R.~W.} \bibnamefont{Chhajlany}},
  \bibinfo{author}{\bibfnamefont{J.}~\bibnamefont{Tuziemski}},
  \bibnamefont{and} \bibinfo{author}{\bibfnamefont{J.}~\bibnamefont{Sperling}},
  \bibinfo{journal}{Phys. Rev. A} \textbf{\bibinfo{volume}{86}},
  \bibinfo{pages}{042123} (\bibinfo{year}{2012}),
  \urlprefix\url{https://link.aps.org/doi/10.1103/PhysRevA.86.042123}.

\bibitem[{\citenamefont{Rana and Parashar}(2012)}]{PhysRevA.85.024102}
\bibinfo{author}{\bibfnamefont{S.}~\bibnamefont{Rana}} \bibnamefont{and}
  \bibinfo{author}{\bibfnamefont{P.}~\bibnamefont{Parashar}},
  \bibinfo{journal}{Phys. Rev. A} \textbf{\bibinfo{volume}{85}},
  \bibinfo{pages}{024102} (\bibinfo{year}{2012}),
  \urlprefix\url{https://link.aps.org/doi/10.1103/PhysRevA.85.024102}.

\bibitem[{\citenamefont{Hassan et~al.}(2012)\citenamefont{Hassan, Lari, and
  Joag}}]{PhysRevA.85.024302}
\bibinfo{author}{\bibfnamefont{A.~S.~M.} \bibnamefont{Hassan}},
  \bibinfo{author}{\bibfnamefont{B.}~\bibnamefont{Lari}}, \bibnamefont{and}
  \bibinfo{author}{\bibfnamefont{P.~S.} \bibnamefont{Joag}},
  \bibinfo{journal}{Phys. Rev. A} \textbf{\bibinfo{volume}{85}},
  \bibinfo{pages}{024302} (\bibinfo{year}{2012}),
  \urlprefix\url{https://link.aps.org/doi/10.1103/PhysRevA.85.024302}.

\bibitem[{\citenamefont{Ali et~al.}(2010)\citenamefont{Ali, Rau, and
  Alber}}]{PhysRevA.81.042105}
\bibinfo{author}{\bibfnamefont{M.}~\bibnamefont{Ali}},
  \bibinfo{author}{\bibfnamefont{A.~R.~P.} \bibnamefont{Rau}},
  \bibnamefont{and} \bibinfo{author}{\bibfnamefont{G.}~\bibnamefont{Alber}},
  \bibinfo{journal}{Phys. Rev. A} \textbf{\bibinfo{volume}{81}},
  \bibinfo{pages}{042105} (\bibinfo{year}{2010}),
  \urlprefix\url{https://link.aps.org/doi/10.1103/PhysRevA.81.042105}.

\bibitem[{\citenamefont{Bhatia}(1986)}]{article1}
\bibinfo{author}{\bibfnamefont{R.}~\bibnamefont{Bhatia}},
  \bibinfo{journal}{Proc Am Math Soc} \textbf{\bibinfo{volume}{96}},
  \bibinfo{pages}{41} (\bibinfo{year}{1986}).

\bibitem[{\citenamefont{Horodecki et~al.}(2003)\citenamefont{Horodecki,
  Horodecki, Horodecki, Horodecki, Oppenheim, Sen(De), and
  Sen}}]{PhysRevLett.90.100402}
\bibinfo{author}{\bibfnamefont{M.}~\bibnamefont{Horodecki}},
  \bibinfo{author}{\bibfnamefont{K.}~\bibnamefont{Horodecki}},
  \bibinfo{author}{\bibfnamefont{P.}~\bibnamefont{Horodecki}},
  \bibinfo{author}{\bibfnamefont{R.}~\bibnamefont{Horodecki}},
  \bibinfo{author}{\bibfnamefont{J.}~\bibnamefont{Oppenheim}},
  \bibinfo{author}{\bibfnamefont{A.}~\bibnamefont{Sen(De)}}, \bibnamefont{and}
  \bibinfo{author}{\bibfnamefont{U.}~\bibnamefont{Sen}},
  \bibinfo{journal}{Phys. Rev. Lett.} \textbf{\bibinfo{volume}{90}},
  \bibinfo{pages}{100402} (\bibinfo{year}{2003}),
  \urlprefix\url{https://link.aps.org/doi/10.1103/PhysRevLett.90.100402}.

\bibitem[{\citenamefont{Coles et~al.}(2019)\citenamefont{Coles, Cerezo, and
  Cincio}}]{PhysRevA.100.022103}
\bibinfo{author}{\bibfnamefont{P.~J.} \bibnamefont{Coles}},
  \bibinfo{author}{\bibfnamefont{M.}~\bibnamefont{Cerezo}}, \bibnamefont{and}
  \bibinfo{author}{\bibfnamefont{L.}~\bibnamefont{Cincio}},
  \bibinfo{journal}{Phys. Rev. A} \textbf{\bibinfo{volume}{100}},
  \bibinfo{pages}{022103} (\bibinfo{year}{2019}),
  \urlprefix\url{https://link.aps.org/doi/10.1103/PhysRevA.100.022103}.

\bibitem[{\citenamefont{Ozawa}(2000)}]{Ozawa_2000}
\bibinfo{author}{\bibfnamefont{M.}~\bibnamefont{Ozawa}},
  \bibinfo{journal}{Phys. Lett. A} \textbf{\bibinfo{volume}{268}},
  \bibinfo{pages}{158} (\bibinfo{year}{2000}),
  \urlprefix\url{https://doi.org/10.1016%2Fs0375-9601%2800%2900171-7}.

\bibitem[{\citenamefont{Debarba et~al.}(2012)\citenamefont{Debarba, Maciel, and
  Vianna}}]{PhysRevA.86.024302}
\bibinfo{author}{\bibfnamefont{T.}~\bibnamefont{Debarba}},
  \bibinfo{author}{\bibfnamefont{T.~O.} \bibnamefont{Maciel}},
  \bibnamefont{and} \bibinfo{author}{\bibfnamefont{R.~O.}
  \bibnamefont{Vianna}}, \bibinfo{journal}{Phys. Rev. A}
  \textbf{\bibinfo{volume}{86}}, \bibinfo{pages}{024302}
  (\bibinfo{year}{2012}),
  \urlprefix\url{https://link.aps.org/doi/10.1103/PhysRevA.86.024302}.

\bibitem[{\citenamefont{Rana}(2013)}]{PhysRevA.87.054301}
\bibinfo{author}{\bibfnamefont{S.}~\bibnamefont{Rana}}, \bibinfo{journal}{Phys.
  Rev. A} \textbf{\bibinfo{volume}{87}}, \bibinfo{pages}{054301}
  (\bibinfo{year}{2013}),
  \urlprefix\url{https://link.aps.org/doi/10.1103/PhysRevA.87.054301}.

\end{thebibliography}


\newpage
\appendix
\section{Supplemental Material.}
We partition the ${\lambda'}^{\downarrow}$ into  two parts as ${\lambda'}^{\downarrow}_{(1)}\bigcup{\lambda'}^{\downarrow}_{(2)}$ with ${\lambda'}^{\downarrow}_{(1)}=\{{\lambda'}^{\downarrow}_{1}, \cdots, {\lambda'}^{\downarrow}_{n}\}$ and ${\lambda'}^{\downarrow}_{(2)}=\{{\lambda'}^{\downarrow}_{n+1}, \cdots, {\lambda'}^{\downarrow}_{N\cdot M}\}$ with $n$ specifying  the minimal  number such that $\tau:=\frac{{\sum^{n}_{i=1}{\lambda'}^{\downarrow}_{i}}-1}{n}\geq{\lambda'}^{\downarrow}_{n+1}$.
The minimum value of $\min_{\lambda_{d}}\|{\lambda'}^{\downarrow}-\lambda^{\downarrow}_{d}\|$  is obtained by taking $\lambda^{\downarrow}_{d}=\lambda^{\downarrow}_{\rm PPT}$ with the element $\lambda^{\downarrow}_{\rm i, PPT}={\lambda'}^{\downarrow}_{i}-\tau$ for $i\leq n$ and $0$ otherwise.  Without loss of generality,  we  assume  that $\lambda^{\downarrow}_{i,PPT} =0$ for $i> N_{+}$  with $N_{+}$ specifying the number of positive elements of ${\lambda'}^{\downarrow}$ as  when $i> N_{+}$ and ${\lambda'}^{\downarrow}_{i}<0$, one has $|{\lambda'}^{\downarrow}_{i}-\lambda_{i,PPT}|>|{\lambda'}^{\downarrow}_{i}|$, namely, subtracting  a positive element leads to an increase in amplitude, and assume that  $\lambda'_{i}\geq \lambda_{i, PPT}$  for $i\geq N_{+}$.   The other elements are given as   ${\lambda'}^{\downarrow}_{i}- \lambda^{\downarrow}_{i,PPT}=\frac{\sum^{n}_{i}{\lambda'}^{\downarrow}_{i}-1}{n}=\tau
$, $ \forall$,  $1\leq i \leq n$.  Otherwise, if another  normalized and positive distribution $\lambda_{\gamma}$ leads to a smaller minimum, for which,   we assume that $\lambda^{\downarrow}_{i, \gamma}=0$ for $i> N_{+}$ and ${\lambda'}^{\downarrow}_{i}-\lambda^{\downarrow}_{i, \gamma}\geq 0$ for $i
\leq N_{+}$ without losing of generality.  As $\|{\lambda'}^{\downarrow}_{i}-{\lambda''}^{\downarrow}_{i}\|^{2}_{2}
={{\lambda'}^{\downarrow}}^{2}_{i}-2\int^{{\lambda''}^{\downarrow}_{i}}_{0}(\lambda'_{i}-x_{i})d x_{i},$   we have
 \begin{widetext}
\begin{eqnarray}
\|{\lambda'}^{\downarrow}-\lambda^{\downarrow}_{\rm PPT}\|^{2}_{2}-\|{\lambda'}^{\downarrow}-\lambda^{\downarrow}_{\gamma}\|^{2}_{2}&=&-2\sum_{j|\lambda^{\downarrow}_{j,\gamma}<\lambda^{\downarrow}_{j,PPT}}\int^{\lambda^{\downarrow}_{j,PPT}}_{\lambda^{\downarrow}_{j,PPT} -\kappa_{j}}
({\lambda'}^{\downarrow}_{j}-x_{j})d x_{j}+
2\sum_{i|\lambda^{\downarrow}_{i,PPT}<\lambda^{\downarrow}_{i,\gamma}}\int^{{\lambda^{\downarrow}}_{i,PPT}+{\delta_{i}}}_{\lambda^{\downarrow}_{i,PPT}}({\lambda'}^{\downarrow}_{i}-y_{i})d y_{i}.\nonumber\\
&\leq & -\tau \sum_{j}\kappa_{j} +\tau\sum_{i}\delta_{i}=0
\end{eqnarray}
 \end{widetext}
where $\delta_{ij}\geq 0$ can always be defined such that,  for the case  $\lambda^{\downarrow}_{i,PPT}\geq \lambda^{\downarrow}_{i, \gamma}$, $\kappa_{i}=\lambda^{\downarrow}_{i,PPT}-\lambda^{\downarrow}_{i, \gamma},$ and for the case  $\lambda^{\downarrow}_{i,PPT}\leq \lambda^{\downarrow}_{  i, \gamma}$, $\delta_{i}=\lambda^{\downarrow}_{ i, \gamma}-\lambda^{\downarrow}_{i,PPT}.$    In the interval of integration,  we have ${\lambda'}^{\downarrow}_{j}-x_{j}\geq {\lambda'}^{\downarrow}_{j}-\lambda_{j, PPT}\geq \tau$ and ${\lambda'}^{\downarrow}_{j}-y_{j}\leq {\lambda'}^{\downarrow}_{j}-\lambda^{\downarrow}_{ j, PPT} \leq \tau, $ and $\sum_{i}\kappa_{i}=\sum_{j}\delta_{j}$. We have completed the proof for  that $\lambda_{\rm PPT}$ is the spectrum achieving the minimum.

To show that $L_{\rm PPT}\geq \frac{\mathcal{N}^{2}}{N_{+}}+\frac{\mathcal{N}^{2}}{N_{-}}$,   we consider  that another kind of   $\lambda^{\downarrow}_{\gamma'}=\lambda^{\downarrow}_{\gamma',(1)}\bigcup\lambda^{\downarrow}_{\gamma',(2)}$ with $\lambda^{\downarrow}_{\gamma',(1)}=\{\lambda^{\downarrow}_{1, \gamma'}, \cdots, \lambda^{\downarrow}_{ N_{+}, \gamma'}\}$ and $\lambda^{\downarrow}_{\gamma', (2)}=\{\lambda^{\downarrow}_{ N_{+}+1, \gamma'}, \cdots, \lambda^{\downarrow}_{ MN, \gamma'}\}$ with $N_{+}$ specifying the number of positive elements in $\lambda'$.
We assume also that $\lambda^{\downarrow}_{i, \gamma'}=0$ for $i>N_{+}$, however,  other elements are not necessary to be positive. This setting, involving looser constraint than that for $\lambda_{d}$, should  lead to a smaller lower-bound.
As  $\min_{\lambda_{\gamma'}}\|{\lambda'}^{\downarrow}_{(1)}-\lambda^{\downarrow}_{\gamma', (1)}\|^{2}_{2}$ is only subject to the constraint that $\sum_{i\leq N_{+}}({\lambda'}^{\downarrow}_{i}-\lambda^{\downarrow}_{i, \gamma'})=\mathcal{N}$ and $2-$norm is convex, the minimum is obtained as $\frac{\mathcal{N}^{2}}{N_{+}}$ when ${\lambda'}^{\downarrow}_{i}-\lambda^{\downarrow}_{i, \gamma'}=\frac{\mathcal{N}}{N_{+}}$  for $1\leq i\leq N_{+}$. Consider the negative terms of ${\lambda'}^{\downarrow}_{(2)}$,  one has $\sum_{i> N_{+}}{\lambda'}^{\downarrow}_{i}=-\mathcal{N}$,    $\|{\lambda'}^{\downarrow}_{(2)}\|\geq \frac{\mathcal{N}^{2}}{N_{-}}$ follows due to the convexity of $2-$ norm, which is saturated when ${\lambda'}^{\downarrow}_{i}=\frac{\mathcal{N}}{N_{-}}$ for $i>N_{+}$ with $N_{-}$ specifying the number of negative terms of $\lambda'$.    Therefore, we have
$$L^{D}_{\rm PPT}\geq \frac{\mathcal{N}^{2}}{N_{+}}+\frac{\mathcal{N}^{2}}{N_{-}}.$$

\end{document}